\documentclass[a4paper]{article}

\usepackage[numbers]{natbib}
\usepackage{authblk}
\usepackage{blindtext}

\begin{document}

\title{Introduction to special issue: Chaos Indicators, Phase Space and Chemical Reaction Dynamics}

\author[1,2]{Matthaios Katsanikas \thanks{mkatsan@academyofathens.gr}}

\author[1,3]{Makrina Agaoglou\thanks{makrina.agaoglou@icmat.es}}

\author[1,4,5]{Francisco Gonz\'alez Montoya\thanks{f.gonzalez.montoya@protonmail.com}}

\affil[1]{School of Mathematics, University of Bristol, Fry Building, Woodland Road, Bristol, BS8 1UG, United Kingdom.}
\affil[2]{Research Center for Astronomy and Applied Mathematics, Academy of Athens, Soranou Efesiou 4, Athens, GR-11527, Greece.}
\affil[3]{Instituto de Ciencias Matem{\'a}ticas, CSIC, C/Nicol{\'a}s Cabrera 15, Campus Cantoblanco, 28049 Madrid, Spain.}
\affil[4]{School of Chemistry, University of Leeds, Leeds, LS2 9JT, United Kingdom.}
\affil[5]{Centro Iternacional de Ciencias AC - UNAM, Avenida Universidad 1001, UAEM, 62210 Cuernavaca, Morelos, Mexico.}

\maketitle

\begin{abstract}
The study of the phase space of multidimensional systems is one of the central open problems in dynamical systems. Being able to distinguish chaoticity from regularity in
nonlinear dynamical systems, as well as to determine
the subspace of the phase space in which instabilities are
expected to occur, is also an important field. To investigate these, diverse chaos indicators have been developed. The information provided from these indicators is important to understand the dynamical behavior of many systems in celestial mechanics and, more recently, in chemical reaction dynamics. This special issue contains new results around the topics: phase space structure, chaos indicators, and chemical reaction dynamics. 
\end{abstract}

\newpage

\section{Introduction}

 The aim of this issue is to present recent developments in phase space analysis, with emphasis on new insights into the connection of phase space structures with transport. These developments are useful for studying the geometry of phase space in Hamiltonian systems with many degrees of freedom and a deeper understanding of the mechanisms of the transport, which is crucial in chemical reaction dynamics, celestial mechanics, and other fields.

A central topic in this special issue is the geometry of the phase space and the methods to study these properties. The methods considered here allow to distinguish chaos from regularity, reveal phase space objects and calculate diffusion properties of the trajectories.
 
 This issue is divided into three research categories with multiple connections between them. These classes are: \textit{Phase Space Structure, Chaos Indicators and Entropy, and Chemical Reaction Dynamics}. In the next sections, we describe each category and underline the basic results of the contributions in this special issue.
 
 This special issue is based on the online workshop ``Chaos Indicators, Phase Space and Chemical Reaction Dynamics'' (4th - 6th May  2020 at Bristol, UK) as a part of the multidisciplinary project Chemistry and Mathematics in Phase Space (CHAMPS)\footnote{CHAMPS is a 6 year EPSRC-sponsored project involving University of Bristol, University of Cardiff, University of Leeds, and Imperial College London. }.  
 
\section{Phase space structure}

The study of the phase space from a geometrical point of view gives us a more complete understanding of the dynamics and transport. The basic geometrical blocks in the phase space are fixed points, periodic orbits, invariant tori, Normally Hyperbolic Invariant Manifolds (NHIMs), and their corresponding invariant manifolds. Calculating and visualising those objects in the high dimensional phase space is an important and challenging problem.

The NHIMs are the generalisation of the hyperbolic periodic orbits in Hamiltonian systems with more than two degrees of freedom \cite{FENICHEL}. Their invariant stable and unstable manifolds could have codimension 1 and direct the dynamics in the multidimensional phase space. The work \cite{GONZALEZ} is a study of a three degrees of freedom open Hamiltonian system close to a partial integrable case. In particular, a numerical procedure to calculate the NHIM and its internal dynamics is presented. With this algorithm, it is possible to study the bifurcations of NHIM when the system is perturbed.

The dividing surface associated with the NHIM defines the flux in the phase space and the associated chemical reaction ratio \cite{WIGNER, Waalkens}. The article \cite{REIFF} shows the use of phase space structures to determine the flux through a region in a time-dependent Hamiltonian system with two degrees of freedom. Also, this work shows how the bifurcations of NHIMs change the reaction probabilities and the connection between NHIMs and reactive trajectories.

In paper \cite{KATSANIKAS}, it is investigated for the first time the structure of periodic orbit dividing surfaces in a non-integrable Hamiltonian system with three degrees of freedom. In this case, the dividing surfaces are 4-dimensional objects in the 5-dimensional phase space. It is also shown that this method can detect homoclinic intersections of the invariant manifolds of the NHIMs. These homoclinic intersections were analysed with the method of Lagrangian descriptors. 

In paper \cite{MIZUNO} of this special issue, an efficient sampling algorithm is proposed for computing reactive islands, which consist of intersections of reactive trajectories and a Poincaré surface of section, and are characterised by the number of times the reactive trajectories intersect the Poincaré surface of section before/after reactions take place. The boundaries of reactive islands are a series of intersections of cylindrical stable and unstable invariant manifolds emanated from a normally hyperbolic invariant manifold located, typically, nearby a rank-one potential energy saddle, which forms ``reactivity boundaries''. The reactive islands structures enable us to predict the fate of trajectories. This method is useful not only for chemical reaction dynamics but in general in nonlinear dynamics.  

In paper \cite{PATSIS}, it is presented the investigation of the evolution of phase space close to complex unstable periodic orbits in two galactic type potentials. It is studied the gradual reshaping of invariant structures close to the transition points (from stability to complex instability and vice versa) and the trace of this in both models. The conclusion is that for time scales significant for the dynamics of galaxies, there are weakly chaotic orbits associated with complex unstable periodic orbits, which should be considered as structure-supporting since they reinforce observed galactic morphological features such as the peanut or X-shaped bulge and the spiral arms.

A new method to find regions of phase space with no KAM tori of a given class is proposed in \cite{KALLINIKOS}. This is applied to the planar circular restricted three-body problem with implications of the location of stable orbits for planets around binary stars. This method will be very useful in many problems of nonlinear dynamics and chemical reaction dynamics.

In this special issue, numerical methods were developed (see \cite{MEISS})  based on the weighted Birkhoff average for studying two-dimensional invariant tori for volume-preserving maps. These methods can compute rotation vectors for regular orbits and distinguish chaos without the use of linearisation or symmetry. Using these methods, resonant and rotational tori are distinguished and critical parameters, where tori are destroyed, can be computed. These were applied in a three-dimensional generalisation of Chirikov’s standard map.

Furthermore, this special issue includes an algorithm that constructs formal (approximate) integrals of motion in time-periodic Hamiltonian systems and the application of this algorithm to three examples (see \cite{TZEMOS}). These integrals of motion depend strongly on the choice of the physical parameters and the initial conditions of the orbits. They can satisfactorily describe the region of central islands of stability, but they often fail to converge in the chaotic zones and close to the region of the escapes.

The work \cite{OZORIODEALMEIDA} develops phase space analysis techniques combined with semiclassical approximations to obtain results concerning the variation of the equilibrium Gibbs ensemble as the temperature ranges from the deep quantum regime to the classical. In particular, it analyses the Gibbs state using its Wigner function, where the Gibbs state is viewed as a density matrix. A doubled phase space is a key part of developing this new semiclassical approach.

Finally, this special issue includes a  new geometric framework of Hamiltonian dynamics (see \cite{DICAIRANO}) that has never been given attention to describe phase space and chaotic dynamics. More precisely, a Hamiltonian flow is identified with a geodesic flow on configuration space–time endowed with a suitable metric due to Eisenhart. This framework has been applied in a Hénon–Heiles model, a two-degrees of freedom system and  1D classical Heisenberg model at a large number of degrees of freedom. A comparison of this geometric framework with other geometric frameworks has been given.

\section{Chaos Indicators and Entropy}

The visualisation of the phase space object and the determination of statistical properties of the trajectories are essential to understand and characterise the dynamics. The chaos indicators are useful tools to visualise objects in phase space and study the transport properties of the trajectories like the diffusion rates, anomalous diffusion, and entropy. 

The article \cite{MOGES} is a study of transport and long term dynamics in multidimensional Hamiltonian systems using the indicators Generalized Alignment Index (GALI) and Maximum Lyapunov Exponent (MLE). Those indicators are based on geometric quantities and are useful to visualise different regions in the phase space with different transport properties. The work compares the differences in properties of the transport in a 2-dimensional map with a multidimensional system constructed with 2-dimensional coupled maps. 

The work \cite{KARMAKAR} is a study of the dynamics of a bipartite molecular system consisting of a trimer coupled weakly to a monomer. Semiclassical insights into the fragmentation dynamics of the condensate are obtained by mapping out the Arnold web of the high dimensional classical phase space. The visualisation of Arnold web in the classical phase space is done using FLI technique. A remarkable correspondence between the quantum state space in occupation number representation and the classical Arnold web in action space is observed.

The delay time is a natural indicator to analyse the phase space of open systems \cite{GONZALEZ}. Using this tool is possible to visualise the tangle between the stable and unstable manifolds of NHIMs in systems with 3 and more degrees of freedom and study the changes in the phase space when the parameters change. In particular, the changes in the stable and unstable manifolds and the possible bifurcations of the NHIMs.

This special issue includes the presentation of the theory and of two examples of Shannon entropy (see \cite{CINCOTTA}) as an efficient indicator that provides a direct measure of the diffusion rate and thus a time-scale for the instabilities arising when dealing with chaos. This indicator provides more dynamical information than a classical chaos indicator.

The indicators are an alternative to studying some properties of high dimensional systems where it is not easy the visualisation of their phase space. The work \cite{SENYANGE} is a study of properties of trajectories of the nonlinear disordered Klein –Gordon lattice chain in one spatial dimension using GALI. This indicator can characterise the behaviour of regular and chaotic trajectories in an efficient way. GALI method can also be efficiently used to study  ``weak'' and ``strong'' chaos.

The complexity of the topological structure of the 2D invariant manifolds is quantified with the topological entropy in \cite{ARENSON} for 3D volume-preserving maps. The structure of the finite segments of the 2D invariant manifolds in the 3D phase space constraints the structure for larger segments of the invariant manifolds. These conditions allow to construct a symbolic dynamics and calculate a lower bound of the topological entropy associated.

Paper \cite{DICAIRANO}, analyses the  Riemannian theory of Hamiltonian chaos. This approach can give us analytical computations of the largest Lyapunov exponent, thus providing at the same time
an explanation of the origin of chaos and a constructive method to compute its strength. One of the important results of this paper about chaos is that the geometrisation of Hamiltonian dynamics
within the framework of the configuration space–time equipped with an Eisenhart metric provides a correct distinction between regular and chaotic motions, and it is in qualitative and quantitative agreement with other geometrisation frameworks.

The theory of Hamiltonian chemical kinetics has been developed in \cite{FARANTOS} based on Hamiltonian thermodynamics in extended phase space and the entropy representation. For dissipating chemical reactions, entropy production is calculated along the associated trajectories. The formalism has been applied to two examples: consecutive first-order elementary chemical reactions taken as a closed thermodynamic system and the nonlinear kinetic equations of an auto-catalytic symmetry breaking chiral model.

\section{Chemical Reaction Dynamics}

The chemical reaction dynamics is at the core of chemistry and its applications. It allows us to determine the chemical reaction ratio using the potentials energy surfaces determined by the electronic structure calculations \cite{SHERRILL}. The chemical reaction ratio is one of the most important observables in the experiments. Some recent examples of construction of potentials can be found in \cite{QCHEM,CHUANG}.

Studying chemical models in time-dependent systems is an important practical problem. Determining the changes of the geometrical objects in the phase space allows one to calculate the change in the reaction ratio in a simple way, without the necessity of large ensembles of trajectories. With this approach, it is possible to efficiently determine the values of the parameters of the system that optimise the chemical ratios. A simple and clear example for time-dependent Hamiltonian systems with 2 degrees of freedom is analysed in \cite{REIFF}.

An important issue of chemical reaction dynamics is the computation of reactive islands because they predict the fate of trajectories. In this special issue, there is a  proposed algorithm (see \cite{MIZUNO}) that estimates regions of reactive islands on the Poincar\'e surface of section as a Voronoi diagram constructed from the set of labelled intersections of computed molecular dynamics trajectories. It is also investigated the efficiency of this algorithm, using a two-degrees of freedom Hamiltonian system of a double-well potential, and further possible extensions of this algorithm are addressed.

An essential step in calculating accurate chemical reactions is the study of intramolecular vibrations. It allows us to understand the different vibrational modes and the redistribution of energy between the reacting molecules before the reaction. The work \cite{KARMAKAR} is a study of the vibrations of a trimer in the presence of a monomer modelled as a Bose-Hubbard four potential well and its classical-quantum correspondence. 
\newpage

Dynamical matching is an important phenomenon in chemical reaction dynamics, and it is encountered in caldera-type potential energy surfaces. This phenomenon was detected for the first time (see \cite{KATSANIKAS}) in a three-dimensional caldera potential energy surface, using periodic orbit dividing surfaces. 

Classical thermodynamics has recently been formulated as a Hamiltonian theory. The work \cite{FARANTOS} shows an extension of the geometric Hamiltonian thermodynamic theory to classical phenomenological chemical kinetics. Using this Hamiltonian representation, it is shown that classical chemical kinetics can also be put in a Hamiltonian framework with Massieu–Gibbs function as the generating Hamiltonian at constant pressure and temperature.

The aim of this issue is not only to expand existing ideas but to reveal how all these can be combined and applied in chemical reactions. We expect this special issue to be the beginning of a new field between chaotic dynamics and Chemical reaction dynamics.
\section{Acknowledgments}

The authors would like to acknowledge the financial support provided by the EPSRC Grant No. EP/P021123/1 ( CHAMPS project). We thank the CHAMPS project and especially the PI Prof. S. R. Wiggins and the project manager E. Machin for their assistance in the realisation of the online workshop ``Chaos Indicators, Phase Space and Chemical Reaction Dynamics'', which was the basis of this special issue. We gratefully thank the Chief Editor V. M. Pérez García and the previous Chief Editor T. Sauer for their support in hosting this collection of articles as a special issue of the journal Physica D: Nonlinear Phenomena. Finally, we thank the production team of Physica D for their help preparing this special issue. 

\bibliographystyle{unsrt}
\bibliography{bibliography}
\end{document}